\documentstyle[12pt,axodraw,psfig]{article}
\def\beq{\begin{equation}}
\def\eeq{\end{equation}}
\def\bea{\begin{eqnarray}}
\def\eea{\end{eqnarray}}
\def\bq{\begin{quote}}
\def\eq{\end{quote}}

\def\gappeq{\mathrel{\rlap
{\raise.5ex\hbox{$>$}}
{\lower.5ex\hbox{$\sim$}}}}
\def\lappeq{\mathrel{\rlap{\raise.5ex\hbox{$<$}}
{\lower.5ex\hbox{$\sim$}}}}
\def\simlt{\stackrel{<}{{}_\sim}}
\def\simgt{\stackrel{>}{{}_\sim}}

\begin{document}
\pagestyle{empty}
\begin{flushright}
{IFT-00/16\\
hep-ph/0008046}
\end{flushright}
\vspace*{5mm}
\begin{center}
{\bf $B^0_{d,s}\rightarrow \mu^-\mu^+$ DECAY IN THE MSSM}
\\
\vspace*{1cm} 
Piotr H. Chankowski, \L ucja S\l awianowska
\\
Institute of Theoretical Physics, Warsaw University, \\
00-681 Warsaw, Poland 
\vskip 1.0cm

\vspace*{1.7cm}  
{\bf Abstract} 
\end{center}
\vspace*{5mm}
\noindent
We present the results of the complete one-loop computation 
of the $B^0_{d,s}\rightarrow l^+l^-$ decay rate in the MSSM. Both 
sources of the FCNC, the CKM matrix and off-diagonal entries of the 
sfermion mass matrices are considered. Strong enhancement of the 
branching ratio (compared to the SM prediction) can be obtained in 
the large $\tan\beta\sim m_t/m_b$ regime in which the neutral Higgs 
boson ``penguin'' diagrams dominate. We make explicit the strong 
dependence of this enhancement on the top squarks mixing angle
in the case of the chargino contribution and on the $\mu$ parameter
in the case of the gluino contribution. We show that, in some regions 
of the MSSM parametre space, the branching ratio for this process can 
be as large as $10^{-(5-4)}$ respecting all existing constraints, 
including the CLEO measurement of $BR(B\rightarrow X_s\gamma)$.
We also estimate, that for chargino and stop masses $\sim{\cal O}(100$ GeV)
$BR(B^0_s\rightarrow l^+l^{\prime -})$ with $ll^\prime=e\tau$ or $\mu\tau$
can be of the order of $10^{-11}$ for the still allowed values of the
off-diagonal entries in the slepton mass matrix.

\vspace*{1.0cm}
 

\rule[.1in]{16.5cm}{.002in}
\vspace*{0.2cm}

\begin{flushleft} 
IFT-00/16\\
August 2000
\end{flushleft}
\vfill\eject
\newpage

\setcounter{page}{1}
\pagestyle{plain}
{\bf  1. Introduction}  

Extensions of the Standard Model (SM) usually predict
new contributions to the flavour changing neutral current (FCNC) 
processes. For example, adding in the most general way a second doublet 
of the Higgs fields to the standard theory of electroweak interactions 
typically leads to large amplitudes of FCNC processes mediated at 
the tree level by neutral Higgs 
particles. Restricting appropriately the possible form of couplings of the 
two Higgs doublets to up- and down-type fermions eliminates such tree 
level contributions to FCNC processes but, of course, new contributions 
induced by loops involving the physical charged scalar still remain. Charged Higgs 
boson contributions to FCNC processes depend however on the same elements of the 
Cabibbo-Kobayashi-Maskawa (CKM) mixing matrix as does the standard $W^\pm$ 
boson contribution and, thus, amplify the effects of the FCNC source 
that is present in the SM, rather than being an independent
new source of such processes. Nevertheless, requiring the effects
of the charged Higgs boson not to spoil succesful predictions of the
standard theory leads to interesting bounds on the 
$(M_{H^+}, \tan\beta)$ plane where $\tan\beta\equiv v_2/v_1$ is the ratio
of the vacuum expectation values of the two Higgs doublets. 
In particular, in the popular two Higgs doublet model of type II (2HDMII),
in which the first doublet couples only to leptons and down-type quarks
and the second one couples to up-type quarks only, processes like 
$B\rightarrow X_s\gamma$ and $K^0\bar K^0$ mixing together with 
$Z^0\rightarrow\bar bb$ 
constrain the plane $(M_{H^+}, \tan\beta)$. In particular, for 
$\tan\beta\simgt3$ $B\rightarrow X_s\gamma$ requires $H^\pm$ to be heavier
than $\sim165-200$ GeV \cite{BOGR}.

Supersymmetric models like the MSSM,  which are the most popular and 
best motivated extensions of the SM, apart from containing a charged
Higgs boson $H^\pm$, induce yet additional contributions to the FCNC 
processes. 
Firstly, in such models the effects of the CKM mixing can be further 
amplified through the loops involving charginos and squarks. Secondly, 
as there is no reason why the squark mass matrices should be diagonal in 
the same (so-called super-CKM) basis as quarks, the sfermion sector of 
such models is in general a new, independent of the CKM matrix, source 
of the FCNC processes. 

Current experimental data on FCNC processes provide important constraints
on these sources of flavour nonconservation in supersymmetric models.
(Extensive reviews are the refs. \cite{GAGAMASI} and \cite{MIPORO}.) 
Taking the CKM matrix as the only source of the FCNC processes, the current
exprimental data on $B\rightarrow X_s\gamma$ and  $K^0\bar K^0$, 
$B^0\bar B^0$ mixings impose some constraints on the MSSM parameter space.
These constraints, which correlate masses and composition of charginos and
top squarks with the mass of the charged Higgs boson, depend in part
on the element $V_{td}$ of the CKM matrix which is not directly measured
\cite{MIPORO,BRFEZW} and become weaker with growing sparticle masses.
Effects of the nonzero off-diagonal entries of the sfermion mass matrices
are usually analyzed separately \cite{GAGAMASI,MIPORO}. 
Stringent constraints apply to the entries causing transitions between
the first two generations. Bounds on the entries connected to the third 
generation, which follow from the $B^0\bar B^0$ mixing and
$B\rightarrow X_s\gamma$ decay are much weaker. Thus, large deviations from the
rates predicted in the SM can still be discovered in the forthcoming
(or already running) experiments like BaBar (SLAC),  BELLE (KEK), CLEO
(Cornell), HERA-B (DESY) and LHC (CERN). In this context, particularily 
interesting process to look at are the decays $B^0_{s(d)}\rightarrow l^+l^-$
because they are clean theoretically being almost free of hadronic 
uncertainties.

Several papers analysed this process in the MSSM 
\cite{BEBOMARI,OTHERS,BAKO} under various assumptions and with different 
approximations. In this paper we perform a complete calculation
\footnote{In what follows we display only formulae for the dominant
contributions. The plots are, however, based on the programme including 
contributions from all relevant one-loop diagrams.} of the process
$B^0_{s,d}\rightarrow l^-l^+$ in the MSSM with emphasise on qualitative 
understanding of the dominant effects. To this end we derive 
simple analytical formulae approximating the main contributions.
We reconfirm that for values of $\tan\beta<20$ the rate of this 
process is not significantly enhanced compared to the prediction of the SM
(apart from the case of $\tan\beta\sim0.5$ and light $H^\pm$ \cite{HENARI}
which is not favoured theoretically within the supersymmetric 
framework). In agreement with earlier papers \cite{OTHERS} we find that 
large enhancement of the branching ratio
is obtained in the case of large $\tan\beta$ values due the neutral 
Higgs boson penguin graphs. This has been previously made explicit  
in ref. \cite{BAKO} (possible role of such contributions to 
$K^0$-$\bar K^0$ and $B^0$-$\bar B^0$ mixings has been emphasized in ref.
\cite{HAPOTO}) in which the contributions of charginos as a source of the 
flavour changing has been considered. We demonstrate strong dependence of
the decay rate on the value of the stop mixing angle and explain it using 
our analytic formulae. Moreover, we extend previous calculations by 
analysing also the case of the flavour mixing induced by squark mass 
matrices. In the latter case we find very strong dependence on the $\mu$ 
parameter. Finally we correlate the predictions for 
$B^0_{s,d}\rightarrow l^-l^+$ with the constraints imposed on the parameter 
space by other processes, in particular by the measurement by CLEO 
\cite{CLEO} of the $B\rightarrow X_s\gamma$ branching ratio.
We find that even respecting all those constraints, 
$BR(B^0_s\rightarrow \mu^-\mu^+)$ can be enhanced up to $10^{-(4-5)}$ for 
$\tan\beta\sim m_t/m_b\approx 50$. Moreover, for such values of $\tan\beta$,
and the off-diagonal 13 entries of the down-type squark mass matrix 
saturating the existing bound \cite{GAGAMASI,MIPORO},
also $BR(B^0_d\rightarrow \mu^-\mu^+)$ can be of the same order of
magnitude. This means that the unsuccesful search done at CLEO \cite{WL}
already provides a constraint on the 13 off-diagonal entries of the down-type
squarks which, for some values of the other MSSM parametres, is stronger 
than the one given in \cite{GAGAMASI,MIPORO}. Finally, we also 
estimate that for chargino and stop masses $\sim{\cal O}(100$ GeV)
$B^0_s\rightarrow l^+l^{\prime -}$ with $ll^\prime=e\tau$ or $\mu\tau$
can be of order $10^{-11}$ for the still allowed values of the
off-diagonal entries in the slepton mass matrix.

{\bf 2. General structure of the amplitude and the SM prediction} 

The effective Lagrangian describing the 
$d_I\bar d_J\rightarrow\bar l_Al_B$ transition has the general form
\begin{eqnarray}
{\cal L}_{eff}= \sum_x C_x {\cal O}_x
\label{eqn:leff}
\end{eqnarray}
in which ${\cal O}_x$ are the local four-fermion operators and $C_x$
are their Wilson coefficients (we suppress quark and lepton flavour 
indices on ${\cal O}_x$ and $C_x$ as well as on various formfactors which
will appear in the following). Four vector-vector operators
${\cal O}^V_{XY}\equiv(\bar d_J\gamma_\mu P_Xd_I)
(\bar l_B\gamma^\mu P_Y l_A)$ and four scalar operators
${\cal O}^S_{XY}\equiv(\bar d_J P_Xd_I)
(\bar l_B P_Yl_A)$ (where $X,Y=LL$, $RR$, $LR$ and $RL$) contribute to this
process. In addition, two tensor operators exist but they do not 
contribute to this process (their matrix elements vanish when taken
between one meson and vacuum states). In the following we will specify the 
formulae to the case of $B^0_{d(s)}=\bar bd(s)$ decay, hence we will take 
$J=3$ and $I=1=d$ for $B^0_d$ or $I=2=s$ for $B^0_s$. Furthemore, 
due to the pseudoscalar nature of the $B^0_I$ mesons we need only two matrix 
elements\footnote{The second follows from the first one by using the QCD 
equation of motion for the quark field operators; 
this fixes their relative sign.}:
\begin{eqnarray}
\langle0|\bar b\gamma^\mu\gamma^5 d_I|B_I^0(q)\rangle&=&-if_{B_I}q^\mu
\nonumber\\
\langle0|\bar b\gamma^5d_I|B_I^0(q)\rangle &=&+if_{B_I}{M^2_B\over m_{d_I}+m_b}
\label{eqn:pcac}
\end{eqnarray}
Using~(\ref{eqn:pcac}) one finds the total $B^0$ width
\begin{eqnarray}
\Gamma = {M_B\over16\pi}f(x^2_A,x^2_B)
\left\{|a|^2\left[1-(x_A-x_B)^2\right]+|b|^2\left[1-(x_A+x_B)^2\right]\right\}
\label{eqn:width}
\end{eqnarray}
where $f(x,y)\equiv\sqrt{1-2(x+y)+(x-y)^2}$, 
$x_A\equiv m_{l_A}/M_B$ and the coefficients $a$ and $b$ are given 
in terms of the Wilson coefficients as
\begin{equation}\label{eqn:a}
a = {f_{B_I}\over4}\left\{(m_{l_B}+m_{l_A}) 
    \left[C^{V}_{LL}-C^{V}_{LR}+C^{V}_{RR}-C^{V}_{RL}\right] 
  - {M_{B_I}^2\over m_b}
    \left[C^{S}_{LL}-C^{S}_{LR}+C^{S}_{RR}-C^{S}_{RL}\right]\right\}
\end{equation}
\begin{equation}\label{eqn:b}
b = -{f_{B_I}\over4}\left\{(m_{l_B}-m_{l_A}) 
     \left[C^V_{LL}+C^V_{LR}-C^V_{RR}-C^V_{RL}\right] 
    -{M_{B_I}^2\over m_b}
     \left[C^S_{LL}+C^S_{LR}-C^S_{RR}-C^S_{RL}\right]\right\}
\end{equation}
where we have neglected $m_{d_I}$ compared to $m_b$.

Three groups of diagrams contribute to the Wilson coefficients:
Box diagrams, $Z^0$ penguin diagrams and neutral Higgs boson penguin diagrams
\footnote{At the one-loop level the photon penguin diagram does not contribute 
for the $\bar ll$ final state due to the vector current conservation. As long as
neutrinos are massless,
the $\bar ll^\prime$ final state can appear neither in the SM nor in the 2HDM;
in the MSSM the $\bar ll^\prime$ final state  can only be due to box 
contribution provided the slepton mass matrices remain non-diagonal in the 
lepton mass eigenstate basis.}.
Denoting the self energy diagrams on the external quark lines as
$-i\Sigma(\not\!p)$, with
\begin{equation}\label{eqn:sig}
\Sigma(\not\!p) = \Sigma_L^V\not\!p P_L + \Sigma_R^V\not\!p P_R 
                + \Sigma_L^SP_L +\Sigma_R^SP_R ,
\end{equation}
and vertex corrections to the couplings
$\bar d_Jd_IZ^0$, $\bar d_Jd_IS^0$ and $\bar d_Jd_IP^0$, where
$S^0(P^0)$ is a neutral scalar (pseudoscalar), respectively as 
\begin{eqnarray}
+i\gamma^\mu(F^V_LP_L+F^V_RP_R) \nonumber\\
-i(F^S_LP_L+F^S_RP_R)\label{eqn:vert}\\
-(F^P_LP_L+F^P_RP_R),\nonumber
\end{eqnarray}
one finds (in the approximation 
$\Sigma(p^2)\equiv\Sigma(0)$, $F(q^2)=F(0)$) the following expressions
for the Wilson coefficients generated by various penguin diagrams: 
\begin{equation}\label{eqn:cvz}
C^V_{XY} = -{e\over 2s_Wc_WM_Z^2}\widehat{F}_X^Vc_Y^e, ~~~~~
X,Y=L,R
\end{equation}
from the $Z^0$ penguin diagram, with $c^e_L=1-2s_W^2$, $c^e_R=-2s^2_W$ and 
$s_W$ ($c_W$) 
is the sine (cosine) of the Weinberg angle; 
\begin{equation}\label{eqn:cvs}
C^S_{LL}=C^S_{LR}=
\sum_{k=1,2}{1\over M_{H_k^0}^2}{Z_R^{1k}\over v_1}\widehat{F}_L^S m_l ~~~~
C^S_{RR}=C^S_{RL}=
\sum_{k=1,2}{1\over M_{H_k^0}^2}{Z_R^{1k}\over v_1}\widehat{F}_R^S m_l
\end{equation}
from neutral scalar penguin diagrams; and 
\begin{equation}\label{eqn:cvp}
C^S_{LL}=-C^S_{LR}=
\sum_{k=1,2}{1\over M_{H_{k+2}^0}^2}{Z_H^{1k}\over v_1}\widehat{F}_L^P m_l~~~~
C^S_{RR}=-C^S_{RL}=
\sum_{k=1,2}{1\over M_{H_{k+2}^0}^2}{Z_H^{1k}\over v_1}\widehat{F}_R^P m_l
\end{equation}
from neutral pseudoscalar penguin diagrams. We use here (and throughout)
the notation of ref.~\cite{RO} in which $H^0_k\equiv(h^0,H^0)$, 
$H^0_{2+k}\equiv(A^0,G^0)$, $H^\pm_k\equiv(H^\pm,G^\pm)$ and $Z_R^{1k}$ 
($Z_H^{1k}$) denotes the projection of the $k$-th physical neutral CP-even 
(-odd) Higgs boson onto the real (imaginary) part of the neutral component 
of the Higgs doublet that couples to the down-type quarks. 
In addition, since at one loop penguin graphs cannot generate transitions 
$B^0\rightarrow\bar ll^\prime$, we have set $m_{l_A}=m_{l_B}=m_l$.
In these formulae
\begin{equation}\label{eqn:formv}
\widehat{F}_{L,R}^V = F_{L,R}^{V} + {e\over2s_Wc_W}c_{L,R}^d\Sigma_{L,R}^V
\end{equation}
where $c^d_L=1-2s^2_W/3$, $c^d_R=-2s^2_W/3$, 
\begin{equation}\label{eqn:forms}
\widehat{F}_{L,R}^S = F_{L,R}^S - {Z_R^{1k}\over v_1}\Sigma_{L,R}^S
\end{equation}
and
\begin{equation} \label{eqn:widehatp}
\widehat{F}_L^P = F_L^P  +{Z_H^{1k}\over v_1}\Sigma_L^P, \quad  
\widehat{F}_R^P = F_R^P - {Z_H^{1k}\over v_1}\Sigma_R^S
\end{equation}
are the full effective vertices including the effects of flavour changing 
self energy diagrams on the external quark lines.
Box diagram contributions to the Wilson coefficients can also be easily found.
From eqs.~(\ref{eqn:a},\ref{eqn:b},\ref{eqn:cvz}-\ref{eqn:cvp}) one sees
that scalar penguin diagrams contribute only to $b$ in eq~(\ref{eqn:width})
whereas the coefficient $a$ receives contributions from both 
$Z^0$ and the pseudoscalar penguin diagrams. The relative sign of the  $Z^0$
and the neutral Goldstone boson contributions to $a$ should be such that the 
total contribution is independent of the gauge chosen for the $Z^0$ propagator.
This is the case if 
\begin{equation}
-m_J\widehat{F}_{L,R}^V + m_I\widehat{F}_{R,L}^V=-M_Z\widehat{F}_{L,R}^P
\nonumber
\end{equation}  
for $P$ referring to the Goldstone boson. Since the formfactor 
$\widehat{F}_{L,R}^P$ for the physical pseudoscalar $A^0$ is related to the 
one for $G^0$ by the $SU_L(2)$ symmetry, this relation tests also the relative 
sign of the $Z^0$ and pseudoscalar penguin diagrams.

The SM contribution to the $B^0\rightarrow ll$ decay is well known
\cite{BUBU} (see also \cite{BUREV}). The Higgs boson couplings to fermions 
are not enhanced so the scalar and pseudoscalar penguins are negligibly small.
The only important box diagram is the one with two $W^\pm$ which contributes
only to $C^V_{LL}$
\begin{equation}
C^V_{LL} =-{1\over16\pi^2}\left({e\over s_W}\right)^4
{\lambda_{tI}\over M_W^2}{x_t\over4}
\left[{1\over 1-x_t} + {\log x_t\over(1-x_t)^2}\right]
\end{equation}        
where  $x_t\equiv(m_t/M_W)^2$ and $\lambda_{tI}\equiv V^\star_{tJ}V_{tI}$. 
The effective $\bar d_Jd_IZ^0$ vertex receives contributions from loops
involving both $W^\pm$ and the charged Goldstone bosons. One finds
\begin{equation}
\widehat{F}_L^V = {1\over16\pi^2}{e^3\over4s^3_Wc_W}
\lambda_{tI} x_t\left[\frac{x_t-6}{1-x_t} -\frac{3x_t+2}{(1-x_t)^2}\log x_t 
\right]
\end{equation}
and $\widehat{F}_R^V=0$ in the limit of $m_{d_I}=0$. Adding all one gets
\begin{equation}\label{eqn:br}
BR(B^0_I\rightarrow\bar ll) = \tau(B^0_I)
\left[{G_F\alpha\over4\pi s_W^2}\right]^2
{f_{\scriptscriptstyle B_I}^2m_l^2M_{\scriptscriptstyle B_I}\over\pi}
|\lambda_{tI}|^2\sqrt{1-4\frac{m_l^2}{M_{\scriptscriptstyle B_I}^2}}
Y^2_0(x_t)
\end{equation}
where $\tau(B^0_I)$ is the lifetime of the $B^0_I$ meson and \cite{BUBUHA,BUREV}
\begin{equation} 
Y_0(x_t)=-{x_t\over8}\left[\frac{x_t-4}{x_t-1} 
+ \frac{3x_t}{(x_t-1)^2}\log x_t  \right]
\end{equation}

In general, taking QCD corrections into account consists of computing 
corrections to the Wilson coefficients at the scale $\sim M_Z$, and 
subsequently evolving 
the latter from the electroweak scale down to the hadronic scale
$\mu_h\sim m_b$. The first step of this procedure amounts to replacing
$Y_0(x_t)$ by  $Y(x_t) = Y_0(x_t) + (\alpha_s/4\pi) Y_1(x_t)$
\cite{BUBUHA} where now $x_t = (\bar m_t(m_t^2)/M_W)^2$.  $Y(x_t)$
can be conveniently parametrized as \cite{LONI}
\begin{eqnarray}
Y(x_t) = 0.997\left[{\bar{m_t}(m_t)\over 166 {\rm GeV}} \right]^{1.55}
\nonumber
\end{eqnarray}
As far as the evolution is concerned, it has been noted in \cite{LONI}
that the vector operators contributing to $B^0\rightarrow\bar ll$ have 
zero anomalous dimensions. Hence, their Wilson coefficients do not 
evolve at all, whereas the evolution of the Wilson coefficients of the 
scalar operators result in multiplying them by $m_b(\mu_h)/m_b(M_Z)$. 
Consequently, if $C^S_{XY}$ are proportional to $m_b(M_Z)$, their 
evolution is taken into account if this factor is replaced by $m_b(\mu_h)$ 
which in turn cancels out with the factor $1/m_b(\mu_h)$ present in 
eqs.~(\ref{eqn:a},\ref{eqn:b}). As it will be apparent, whenever the 
coefficients $C^S_{XY}$ are large, they are indeed proportional to 
$m_b(M_Z)$. In the SM, including QCD corrections one finds \cite{LONI}
\begin{equation}
BR(B^0_s\rightarrow\bar \mu\mu) = 4.1\times 10^{-9}
\left[{\tau(B_s)\over1.54 ~ps}\right]
\left[{f_{\scriptscriptstyle B_s}\over245 {\rm MeV}}\right]^2
\left[{|V_{ts}|\over0.040}\right]^2
\left[{\bar m_t(m_t)\over166 {\rm GeV}}\right]^{3.12}
\end{equation}

{\bf 3. Contribution of the extended Higgs sector}

As remarked in the introduction, the presence of the physical charged 
Higgs boson in the extended Higgs sector of the MSSM (or 2HDM) in general
enhances the FCNC transition rates generated by the CKM mixing matrix. This 
enhancement can appear through the $H^\pm$ contribution to box diagrams, $Z^0$ 
penguin diagrams and neutral Higgs boson penguin diagrams. The latter type of 
diagrams can only be important in the large $\tan\beta\simgt30$ regime in which 
the neutral Higgs boson couplings to the down-type quarks and charged leptons 
are enhanced by $\tan\beta$ factors.

For low values of $\tan\beta\simlt20$ the neutral Higgs boson penguin 
diagrams are small. It is also easy to check, that for such $\tan\beta$ 
values no box diagram can give significant contribution. Thus, the only 
large contribution can be due to the $H^\pm$ contribution to the $Z^0$ 
penguin diagrams. Computing the relevant self energy diagrams (vector 
parts thereof) and vertex corrections one arrives at
\begin{eqnarray}
\Delta\widehat F_L^V = {1\over16\pi^2}{e^3\over(s_Wc_W)^3} 
\lambda_{tI} \cot^2\beta{m_t^2\over M_Z^2}{1\over4} 
\frac{y_t}{1-y_t}\left[1 +\frac{1}{1-y_t}\log y_t\right]\nonumber\\
\Delta\widehat F_R^V = -{1\over16\pi^2}{e^3\over(s_Wc_W)^3}
\lambda_{tI} \tan^2\beta {m_bm_{d_I}\over M_Z^2}{1\over4} 
\frac{y_t}{1-y_t}\left[1 +\frac{1}{1-y_t}\log y_t\right]\nonumber
\end{eqnarray}
where $y_t\equiv(m_t/M_{H^+})^2$. Taking into account 
only $\Delta\widehat F_L^V$ which is enhanced for $\tan\beta<1$ 
amounts to replacing $Y(x_t)$ in eq.~(\ref{eqn:br}) by
\begin{equation}
Y(x_t)\rightarrow Y(x_t) - \cot^2\beta {x_t\over8} 
\frac{y_t}{1 - y_t} \left[1 + \frac{1}{1 - y_t}\log y_t\right] 
\end{equation}
The new contribution has the same sign as $Y(x_t)$ and, therefore,
enhances the SM contribution. For example, for $\tan\beta=0.5$ and
$M_{H^+}=M_W$, $BR(B^0\rightarrow\bar ll)$ is enhanced by a factor of
$\left( 1 + \frac{1.566}{0.997}\right)^2 \approx 6.6$ compared to the
SM prediction. 

For large $\tan\beta\sim m_t/m_b$ in the case of $B^0_s$ decay,
$\widehat F^V_R$, despite being  suppressed by one
power of $m_s/M_W$, is about two orders of magnitude larger than $\widehat F^V_L$ 
and, for $M_{H^+}\sim100$ GeV, is of the order of the SM contribution. 
However, in this regime, there are other contributions which are
more important \cite{SKKA,LONI}. 

Firstly, the mixed, $W^\pm H^\pm$, 
box diagram in which $H^\pm$ couples to the $b$-quark is also
${\cal O}(\tan^2\beta)$ and is not suppressed by $m_{d_I}/M_W$. After summation
over different types of virtual quarks it gives \cite{LONI}
\begin{equation}
C_{LR}^S = {1\over16\pi^2}\left(\frac{e}{s_W}\right)^4
\frac{m_l m_b}{M_W^2} \lambda_{tI}\tan^2\beta{1\over4}
\frac{x_t}{x_H-x_t}\left[\frac{1}{x_H-1}\log x_H - \frac{1}{x_t-1}\log x_t
\right] 
\end{equation}
where $x_H\equiv(M_{H^\pm}/M_W)^2$. The other $W^\pm H^\pm$ box is 
proportional to $m_{d_I}/M_W$ and hence it is less important. The box diagrams
containing two chaged scalars (either physical or Goldstone) are suppressed
always by $(m_l/M_W)^2$. Therefore, although the $H^\pm H^\pm$ box
grows as $\tan^4\beta$, it is not important even for $\tan\beta\sim50$.

Secondly, there are neutral Higgs boson penguin diagrams. 
It turns out \cite{LONI}, that the dominant i.e. $\sim\tan\beta$ part of 
the genuine $\bar d_Jd_IS^0(P^0)$ vertex correction cancels out
\footnote{This cancellation is even simpler in the case of the MSSM 
than in the case of the 2HDM(II) considered in \cite{LONI}.} and the 
only contribution arises from the scalar parts of the self energies of
external quarks (for $\bar B^0$ decay it is $\Sigma^S_R$ which is dominant):
\begin{equation}
\Sigma^S_L = {1\over16\pi^2}\left(\frac{e}{s_W}\right)^2 m_b\lambda_{tI} 
x_t\frac{x_H - 1}{x_H - x_t}\left[\frac{x_H}{x_{H}-1}\log x_H -
\frac{x_{t}}{x_t-1} \log x_t \right]
\end{equation}
where $x_H = (M_{H^\pm}/M_W)^2$. 

\begin{figure}
\psfig{figure=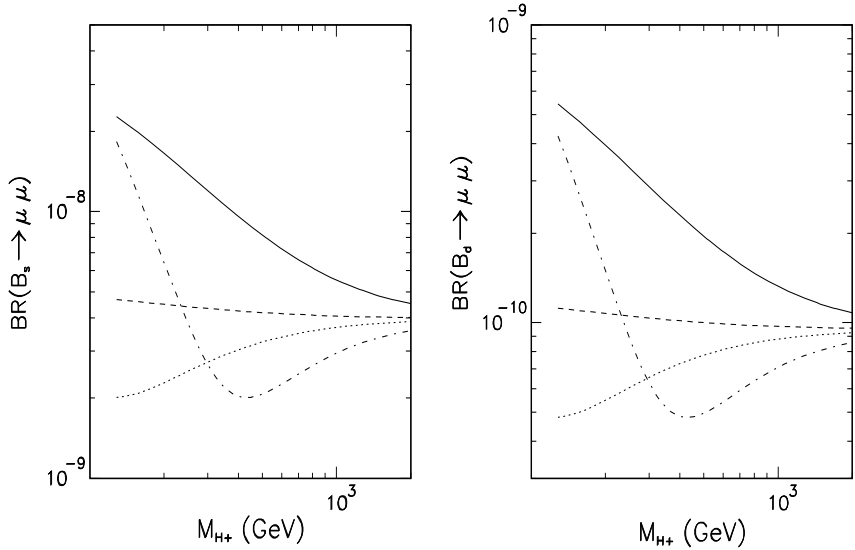,width=12.0cm,height=8.cm} \vspace{1.0truecm}
\caption{Contribution of the Higgs sector of the MSSM or 2HDM(II) to 
$BR(B^0_{s(d)}\rightarrow\mu^-\mu^+)$ as a function of the charged Higgs
boson mass for $\tan\beta=$ 0.5 (solid lines), 2 (dashed), 25 (dotted)
and 50 (dot-dashed).}
\label{fig:bll1}
\end{figure}

Using the 
formulae~(\ref{eqn:formv},\ref{eqn:widehatp}) and the fact that in the MSSM
for neutral CP-even scalars for large values of $\tan\beta$ the following 
relations hold\footnote{The well-known large radiative corrections
to $h^0$ and $H^0$ masses do not spoil these relations. Moreover, 
these corrections do not affect the neutral Higgs boson penguin contributions
because they always modify significantly only the mass of that Higgs boson 
which almost does not couple to the down-type quarks and charged leptons.}
\begin{eqnarray}
\sin^2\alpha\approx1,\quad M_h^2 \approx M_A^2\quad 
{\rm for} \quad M_A < M_Z\nonumber\\ 
\cos^2\alpha\approx1,\quad M_H^2 \approx M_A^2\quad 
{\rm for} \quad M_A > M_Z\\ 
M^2_{H^+}=M^2_A + M^2_W\phantom{aaaaaaaaaa}\nonumber
\end{eqnarray}
we can now summarize the dominant contribution of the extended Higgs 
sector to the the coefficients
$a$ and $b$ given by eqs.~(\ref{eqn:a},\ref{eqn:b}) \cite{LONI}:
\begin{eqnarray}
a&=&{1\over16\pi^2}{f_{B_I}\over2}\left(\frac{e}{s_W}\right)^4
\frac{m_l}{M_W^2}\lambda_{tI} 
\left[Y(x_t) - \frac{M_{B_I}^2}{8M_W^2}\tan^2\beta\frac{\log r}{r-1}\right]
\nonumber\\
b&=& -{1\over16\pi^2}{f_{B_I}\over2}\left(\frac{e}{s_W}\right)^4
\frac{m_l}{M_W^2}\lambda_{tI} 
\frac{M_{B_I}^2}{8M_W^2}\tan^2\beta\frac{\log r}{r-1}
\end{eqnarray}
where $r\equiv1/y_t=(M_{H^+}/m_t)^2$.
Since $\frac{\log r}{r-1}>1$, the CP-odd neutral Higgs exchange interferes
destructively with the SM contribution.
Figures~\ref{fig:bll1} show the contribution of the extended Higgs sector
of the MSSM (assuming that sparticles contribute negligibly) 
or of the 2HDM(II)\footnote{In the case of the 2HDM(II) the subleading in
$\tan\beta$
contributions of the genuine vertex corrections in the CP-even Higgs boson 
penguin may be different than in the MSSM because the dimensionfull couplings
$H^+H^-H^0(h^0)$ differ in both models \cite{SKKA}. Still, unless these
coulings in the 2HDM(II) are very large (and numerically very different
from their MSSM counterparts) so as to enhance the otherwise
subleading contribution, figs~\ref{fig:bll1} should be fairly representative
also for the 2HDM(II) results.}
to $BR(B^0_{s,d}\rightarrow\mu^-\mu^+)$ as a function of $M_{H^+}$ for 
different values of $\tan\beta$. These result which should be compared
with the SM results $4\times10^{-9}$ and $1\times 10^{-10}$, respectively,
agree for $\tan\beta\simgt30$ with the ones given in \cite{LONI} and,
for smaller values of $\tan\beta$,
update the computations done earlier in refs. \cite{HENARI,SKKA}.

{\bf 4. Chargino contribution} 

Another source of amplification of the flavour changing transitions
induced by the CKM matrix is the chargino sector of the MSSM. Assuming 
that the squark mass matrices are diagonal in the super-CKM basis,
the first result is that in the whole relevant parameter space the box
diagram contribution to any of the Wilson coefficients remains small
compared to the SM contribution. Furthermore, the $Z^0$ penguin can 
change the predicted $BR(B^0_{s,d}\rightarrow\mu^-\mu^+)$ by no more
than $\sim$5-10\% for $\tan\beta\sim2$ and $\sim$20\% for $\tan\beta\sim0.5$.
The magnitude and sign of this contribution depends, apart from the masses 
of the sparticles involved, also on the chargino composition and on the
mixing angle of the top squarks. 
For natural stop composition i.e. when the lighter stop is 
predominantly right-handed and the mixing angle is not too large
\cite{CHPOsusy}, the chargino loop contribution to the Wilson coeffcients 
has opposite sign to that of the top quark 
loop and, hence, decreases the rate of the $B^0_{s,d}\rightarrow\mu^-\mu^+$
decay. This is very similar to the opposite, as compared to the SM, sign 
of the chargino-stop loop contribution to 
$R_b\equiv\Gamma(Z^0\rightarrow\bar bb)/\Gamma(Z^0\rightarrow hadr)$ 
\cite{CHPORb} since, in view of the smallness of the box
contribution, the two calulations are very similar
We conclude that in the whole range of the MSSM 
parameter space the box and $Z^0$-penguin diagrams arising from chargino
exchanges do not change the order of magnitude of the 
$B^0_{s,d}\rightarrow\mu^-\mu^+$ decay rate.

\begin{figure}[htbp]    
\begin{center}
\begin{picture}(400,150)(0,0)
\Line(60,30)(120,30)
\Line(87,33)(93,27)
\Line(87,27)(93,33)
\ArrowLine(0,30)(60,30)
\ArrowLine(180,30)(120,30)
\Text(10,25)[t]{$q_B$}
\Text(170,25)[t]{$d^c_A$}
\Text(70,25)[t]{$\tilde{H}_{u}$}
\Text(75,50)[t]{$U^{c}_K$}
\Text(110,50)[t]{$Q_L$}
\Text(110,25)[t]{$\tilde{H}_{d}$}
\Text(70,130)[t]{$H_{u}^{*}$}
\DashArrowLine(90,90)(50,30){4}
\DashArrowLine(90,90)(130,30){4}
\DashArrowLine(90,90)(90,140){4}
\Text(90,0)[t]{$a)$}

\Line(260,30)(320,30)
\Line(287,33)(293,27)
\Line(287,27)(293,33)
\ArrowLine(200,30)(260,30)
\ArrowLine(380,30)(320,30)
\Text(210,25)[t]{$q_B$}
\Text(370,25)[t]{$d^c_A$}
\Text(270,25)[t]{$\tilde{H}_{u}$}
\Text(275,50)[t]{$U^{c}_K$}
\Text(310,50)[t]{$Q_L$}
\Text(310,25)[t]{$\tilde{H}_{d}$}
\Text(270,130)[t]{$H_{d}$}
\DashArrowLine(290,90)(250,30){4}
\DashArrowLine(290,90)(330,30){4}
\DashArrowLine(290,140)(290,90){4}
\Text(290,0)[t]{$b)$}\end{picture}
\end{center}
\caption{Diagrams giving rise to $\Delta_u Y_d$ i~$\Delta_d Y_d$ 
respectively, in the construction of the effective theory.}
\label{fig:bll2}
\end{figure}
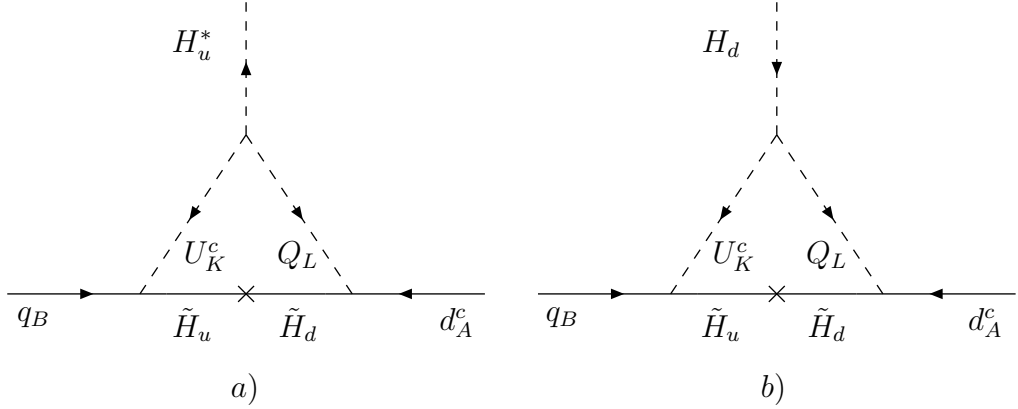

Huge contribution to this rate can be however induced for large
$\tan\beta\simgt30$ by neutral Higgs boson penguins. This has been first
maded explicit in ref. \cite{BAKO} in the approach based on the effective 
lagrangian method. Our numerical results are obtained by full computation 
of all relevant Feynman diagrams. Here we present only the derivation
of the approximate formulae summarizing the dominant effects. To this end
we consider the limit in which all soft SUSY breaking parameters, except
for the ones which determine the Higgs potential, are much larger than the 
electroweak scale. In this limit, which allows us to work in the symmetric 
phase of the theory (i.e. with $v_i=0$) in which sfermions still have 
definite chirality, we can construct the effective theory by integrating out 
sparticles (but not the Higgs fields). In this construction, threshold corrections
shown in fig.~\ref{fig:bll2} give rise to the effective Yukawa interactions
of the down-type quarks summarized by
\begin{eqnarray} \label{eqn:eff}
{\cal L}_{eff}=-\; \epsilon_{ij}
\left(Y_d + \Delta_d Y_d\right)^{BA} H_i^dq^A_jd^{c\;B} 
-\left(\Delta_u Y_d\right)^{BA}H_i^{*u} q^A_i  d^{c\;B} + h.c.
\end{eqnarray}
where $A$, $B$ are the generation indices and we work in the language of 
two-component Weyl spinors. In order to diagonalize the quark mass matrix
arising after the electroweak symmetry breaking we perform first the 
standard CKM rotations (diagonalizing the original matrix $Y_d^{BA}$)
followed by the infinitesimal rotations
\begin{equation} \label{eqn:rot}
d^A \to \left(1 + \Delta V_L^{D \dagger}\right)^{AB}d^B, 
\quad d^{c\;A} \to d^{c\;B}\left(1 + \Delta V_R^{D \dagger}\right)^{AB}
\end{equation}
with  $\Delta V_L^{D \dag}$, $\Delta V_R^{D \dag}$ satisfying
$\Delta V_{L,R}^{D \dag} = -\Delta V_{L,R}^{D}$. Diagonal mass matrix
for down-type quarks is obtained with
\begin{equation}\label{eqn:mm}
-\left(\Delta^\prime_d Y_d\right)^{AB} + 
\frac{v_2}{v_1}\left(\Delta^\prime_u Y_d\right)^{AB} =
(\Delta V_R^{D \dagger})^{AB}Y_d^B + Y_d^A (\Delta V_L^{D})^{AB} 
\end{equation}
where $Y_d^A$ are already diagonal and $\Delta^\prime_{u(d)} Y_d$ are 
related to the original $\Delta_{u(d)}Y_d$ by the rotation diagonalizing 
the original $Y_d^{BA}$. This leads
\footnote{Strictly speaking eq.~(\ref{eqn:mm}) must hold only for 
off diagonal elements; for $A=B$ the relation $m_{d_A}=-Y^A_dv_1/\sqrt2$
is corrected but the net result is that in eq.~(\ref{eqn:effyuk})
$Y^A_d \equiv -\sqrt2 m_{d_A}/v_1$ again.}
to the effective Yukawa couplings of the neutral Higgs bosons of the form
\begin{eqnarray} 
{\cal L} = -{1\over\sqrt2} d^c\left(-Y_d Z_R^{1k} + 
\Delta^\prime_u Y_d Z_R^{2k}  - \tan\beta\Delta^\prime_u Y_d Z_R^{1k} 
\right)d \;H^0_k + h.c.\nonumber\\ 
+ {i\over\sqrt2}d^c\left(Y_d Z_H^{1k}  + \Delta^\prime_u Y_d Z_H^{2k} 
+ \tan\beta\Delta^\prime_u Y_d Z_H^{1k} \right)d \;H^0_{k+2} + h.c.
\label{eqn:effyuk}
\end{eqnarray} 
(In the lagrangian~(\ref{eqn:effyuk}) $Y_d$ is diagonal and the 
rest of the notation is explained below eq.~(\ref{eqn:cvp})) which in
general generates the FCNC transitions. Note that the correction 
$\Delta_d Y_d$ dissapeared altogether as it should, since it cannot 
cotribute to the FCNC transition. 

The correction $\Delta_u Y_d$ in eq.~(\ref{eqn:eff}) is easily computed 
in the basis in which $Y^{AB}_u$ is diagonal and the initial 
$Y^{AB}_d=Y^A_dV^\dagger_{AB}$ where $V_{AB}$ is the CKM matrix.
Starting from the SUSY breaking part of the lagrangian \cite{RO}
\begin{eqnarray}
{\cal L}_{soft} = 
- \left(m_U^2\right)^{AB}\;U^{c*A} U^{cB}\; 
- \left(m_Q^2\right)^{AB}\;Q^{*A}_i Q^B_j
+ \left(\epsilon_{ij}A_U^{AB} H_u^i Q^A_j U^{cB}\; 
- \mu\epsilon_{ij}\tilde H^i_d \tilde H^j_u + h.c.\right)
\nonumber
\end{eqnarray}
we obtain
\begin{equation}\label{eqn:duyd}
\Delta_u Y_d^{AB} = {1\over16\pi^2}Y_d^{AC}A_U^{BC}Y_u^B 
\mu C_0(\mu^2, M_{Q_B}^2, M_{U^c_C}^2) 
\end{equation}
where $C_0$ is the standard three-point function. 
\begin{equation}
C_0(a,b,c)={1\over a-b}
\left[{a\over a-c}\log{a\over c}-{b\over b-c}\log{b\over c}\right]
\end{equation}
Inserting (\ref{eqn:duyd}) in eq.~(\ref{eqn:eff}) and performing all steps
making the usual assumption $A_U^{AB} = Y_u^A A_u^A \delta^{AB}$ (i.e. that the 
trilinear soft terms are proportional to the Yukawa couplings)
and keeping only  the top Yukawa coupling  leads to
\begin{equation}\label{eqn:dy}
\Delta^\prime_u Y_d^{JI} = \pm{1\over16\pi^2}\lambda_{tI}Y_d^J
Y_t^2 A_t m_{C_1}  C_0(m_{C_1}^2, M_{\tilde{t}_L}^2, M_{\tilde{t}_R}^2) 
\end{equation}
where we have replaced $\mu$ with $m_{C_1}$ and the sign $\pm$ keeps
track of the sign of $\mu$. Using eq.~(\ref{eqn:effyuk}) in 
eqs.~(\ref{eqn:cvs},\ref{eqn:cvp}) yields the full vertex formfactors
\begin{eqnarray}
\widehat{F}^S_L = 
{1\over\sqrt2}\Delta^\prime_u Y_d \left[Z_R^{2k} - Z_R^{1k}\tan\beta\right]
\approx-{1\over\sqrt2}\Delta^\prime_u Y_d Z_R^{1k}\tan\beta\nonumber\\ 
\widehat{F}^P_L = 
{1\over\sqrt2}\Delta^\prime_u Y_d \left[Z_H^{2k} + Z_H^{1k}\tan\beta\right]
\approx{1\over\sqrt2}\Delta^\prime_u Y_d Z_H^{1k}\tan\beta
\label{eqn:formfs}
\end{eqnarray}
(right formfactors are given by the Hermitean conjugation; they involve 
$Y_d^I$ and are, therefore, subleading).

\begin{figure}
\psfig{figure=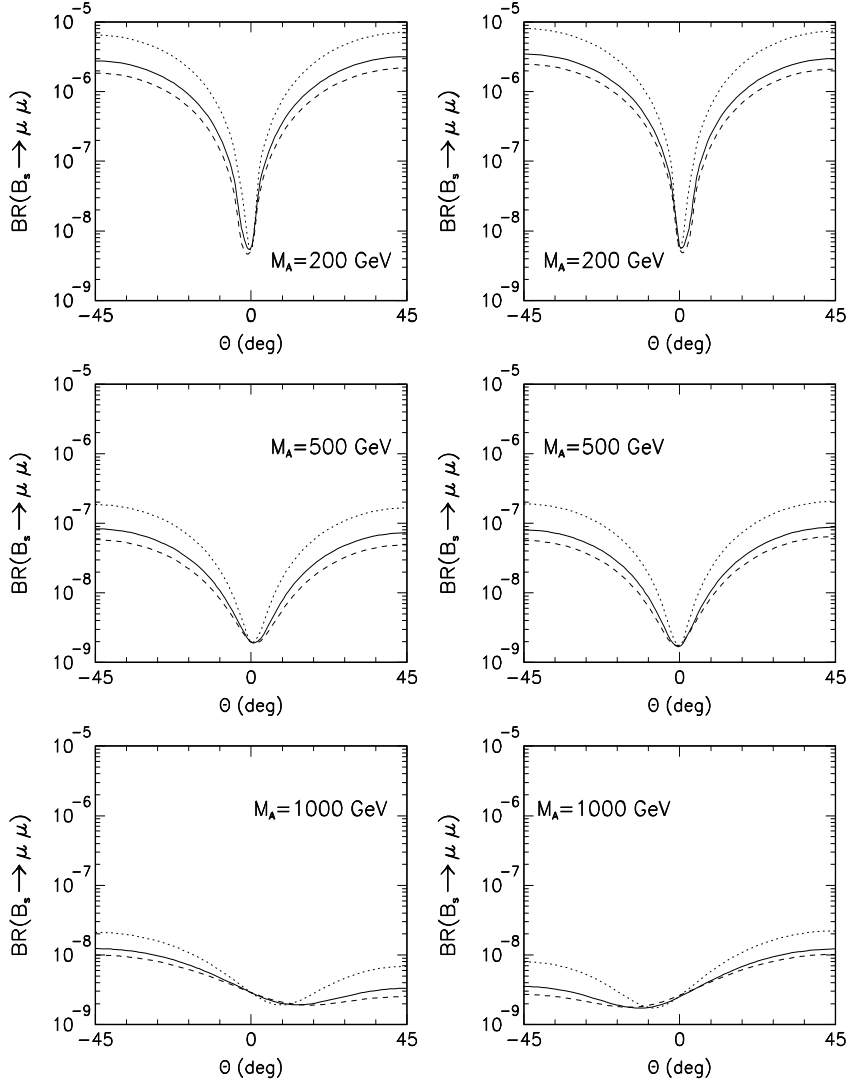,width=12.0cm,height=15.cm} \vspace{1.0truecm}
\caption{$BR(B^0_{s(d)}\rightarrow\mu^-\mu^+)$ as a function of the 
stop mixing angle $\theta_t$ for $\tan\beta=50$, the lighter chargino 
mass 100 GeV and different values of $M_A$. Solid, dashed and dotted
lines correspond to $(M_{\tilde t_2},M_{\tilde t_1})$ equal
(240,500), (400, 700) and (300, 850) GeV, respectively. In the
left (right) pannels $M_2/\mu=10(-1)$, where $M_2$ is the usual $SU(2)$
gaugino mass parameter.}
\label{fig:bll3}
\end{figure}

Detailed comparison of the above simplified calculation with the 
standard diagramatic approach (in which one computes both, the self energy 
corrections the 1PI vertex diagrams, in the phase in which the electroweak 
symmetry is broken) reveals that the dominant contributions given by 
eqs.~(\ref{eqn:formfs}) to the formfactors arise only from the self energy 
diagrams (the 1PI vertex corrections contain one power of $\tan\beta$ less). 
Moreover, the comparison shows that one should replace $A_t$ 
by $\tilde A_t\equiv A_t+\mu\cot\beta$, $M_{\tilde t_L}$, $M_{\tilde t_R}$ 
with the true mass eigenstates $M_{\tilde t_1}$, $M_{\tilde t_2}$
and justifies the replacement of $\pm\mu$ by the mass of the lighter chargino.

Using eqs.~(\ref{eqn:formfs}) we get 
\begin{equation}\label{eqn:a_charg}
a = {f_B\over4}{1\over16\pi^2}\lambda_{tI}  \frac{2m_l}{M_W^2} 
\left(\frac{e}{s_W}\right)^4 
\left[Y(x_t) - \frac{M_B^2}{8M_W^2}\tan^2\beta\frac{\log r}{r-1} \pm  
\frac{M_B^2}{8M_W^2}\frac{m_t^2}{M^2_{A}}
\tan^3\beta\tilde A_t m_{C_1} C_0 \right] 
\end{equation} 
\begin{equation}\label{eqn:b_charg}
b = {f_B\over4}{1\over16\pi^2}\lambda_{tI} \frac{2m_l}{M_W^2}
\left(\frac{e}{s_W}\right)^4\left[-\frac{M_B^2}{8M_W}
\tan^2\beta\frac{\log r}{r-1}\pm\frac{M_B^2}{8M_W^2} 
\frac{m_t^2}{M^2_{A}}\tan^3\beta\tilde A_t m_{C_1} C_0 \right]
\end{equation} 
Knowing that $Y(x_t)\approx1$ these formulae allow for a quick estimate of the
effects. It is important to note that the contribution of chaginos
to the rate grows as $\tan^6\beta$ and therefore can be  much larger than
the contribution of the Higgs sector.

\begin{figure}
\psfig{figure=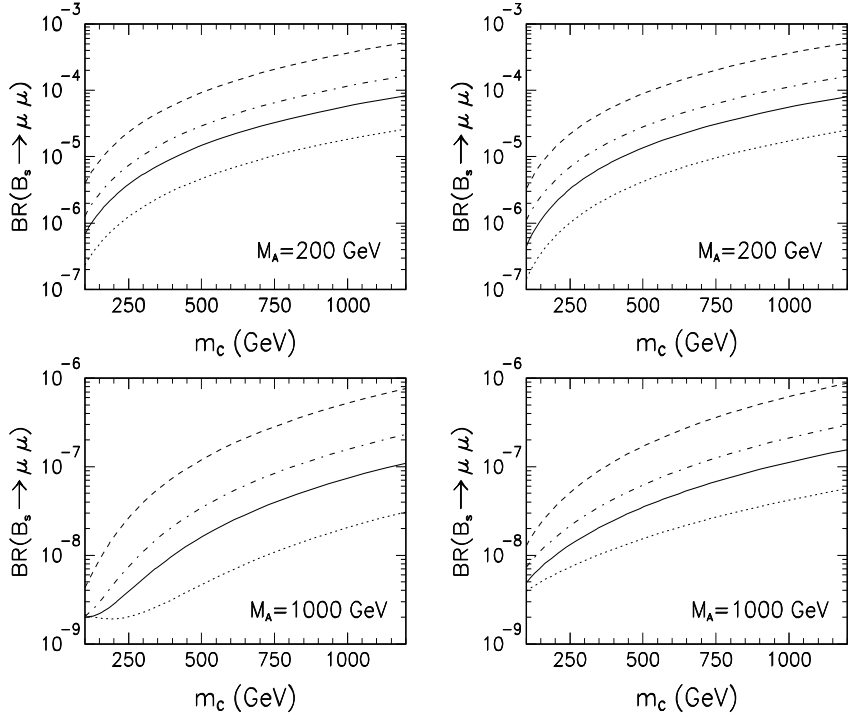,width=12.0cm,height=11.cm} \vspace{1.0truecm}
\caption{$BR(B^0_{s(d)}\rightarrow\mu^-\mu^+)$ as a function of the 
lighter chargino mass for $\tan\beta=50$, $M_A=200$ and 1000 GeV.
Solid (dashed) lines correspond to $(M_{\tilde t_2},M_{\tilde t_1})$ equal
to $(m_{C_1},3m_{C_1})$ and the stop mixing angle $\theta_t=10^o$ ($30^o$)
whereas dotted (dash-dotted) lines to $(3m_{C_1},5m_{C_1})$ 
and $\theta_t=10^o$ ($30^o$), respectively. In the
left (right) pannels $M_2/\mu=10(-1)$.}
\label{fig:bll4}
\end{figure}

Figure~\ref{fig:bll3} shows the dependence of the full branching ratio
$BR(B^0_s\rightarrow\mu^-\mu^+)$, including the SM, Higgs boson and 
chargino contributions, on the mixing angle of the top squarks for
some values of the other MSSM parameters. The minimum around 
$\theta_t\approx0$ corresponding to $\tilde A_t\approx0$ is clearly seen. 
Incidentally this plot also supports the replacement of $\mu$ by $\pm m_{C_1}$
in eq.~(\ref{eqn:dy}) because very similarly (up to a reflection
$\theta_t\rightarrow-\theta_t$ which follows from different signs of $\mu$)
looking curves in the left 
and right panels have the same $m_{C_1}$ but distinctly different $\mu$.
Another important feature of the chargino contribution is that it does 
not vanish if all sparticle mass parameters are scaled uniformly:
$M_{\tilde{t}_i} \rightarrow \lambda M_{\tilde{t}_i}$, 
$m_{C_i}\rightarrow\lambda m_{C_i}$, 
$\mu\rightarrow\lambda\mu$, $A_t\rightarrow \lambda A_t$.
This is clear from the fact that in such a case $C_0\rightarrow\lambda^{-2}C_0$.
This is illustrated in figure~\ref{fig:bll4} which shows 
$BR(B^0_s\rightarrow\mu^-\mu^+)$ as a function of the lighter chargino mass
for $(M_{\tilde{t}_2},M_{\tilde{t}_1})$ equal to $(m_C,3 m_C)$ 
and $\theta_t =10^o$ (solid lines), $(m_C,3 m_C)$ 
and $\theta_t =30^o$ (dashed lines), $(3m_C,5 m_C)$ 
and $\theta_t =10^o$ (dotted lines) and $(3m_C,5 m_C)$ 
and $\theta_t =30^o$ (dash-dotted lines). 
In fact, keeping the stop mixing angle fixed requires that $\tilde A_t$ 
scales as $\lambda^2$ rather than as $\lambda$, which explains the growth 
of the rates with $m_{C_1}$ in fig.~\ref{fig:bll4}.

\begin{figure}
\psfig{figure=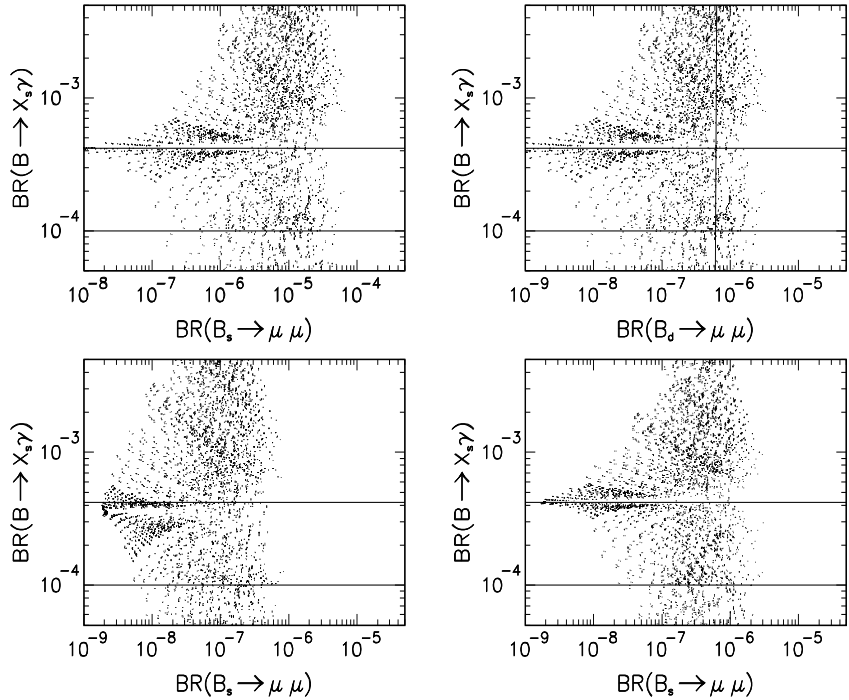,width=12.0cm,height=11.cm} \vspace{1.0truecm}
\caption{$BR(B\rightarrow X_s\gamma)$ versus
$BR(B^0_{s(d)}\rightarrow\mu^-\mu^+)$ for 
$\tan\beta=50$, $M_A=200$ GeV in panels a) and b), $\tan\beta=50$, $M_A=600$ 
GeV in panel c) and $\tan\beta=30$, $M_A=200$ in panel d). Limits from CLEO
on $BR(B\rightarrow X_s\gamma)$ and on 
$BR(B^0_{d}\rightarrow\mu^-\mu^+)$ are also shown by solid lines.}
\label{fig:bll5}
\end{figure}

To check the correlation of the prediction for 
$BR(B^0_{s(d)}\rightarrow\mu^-\mu^+)$  with the reslts for 
$BR(B\rightarrow X_s\gamma)$ we have performed scans over the relevant 
parameter space of the MSSM. We took the following ranges:
$100< m_{C_1}<1000$ GeV, $0.1<|M_2/\mu|<10$, $1<M_{\tilde t_2}/m_{C_1}<10$, 
$1<M_{\tilde t_1}/M_{\tilde t_2}<5$ and $-60^o<\theta_t<60^o$ and rejected 
points for which $\Delta\rho_{squarks}>6\times10^{-4}$ and $M_h<107$ GeV.
For calculating $BR(B\rightarrow X_s\gamma)$ we have used 
the routine based on refs. \cite{MIK,NEU} including the NLO matching 
conditions at the scale $M_Z$ for the top and charged Higgs contribution as in
\cite{ADYA,CIDEGAGI} and only the LO ones for the supersymmetric contribution
\cite{BAGI,BEBOMARI}. We have not used the available NLO matching conditions
for the supersymmetric particles since they are computed under the specific
assumptions about the sparticle spectrum, not necessarily satisfied in the 
scan and, moreover, they are not valid
for large values of $\tan\beta$. The theoretical 
uncertainty is taken into account by computing the rate for $\mu_h=2.4$ and 
9.6 GeV and then by shifting its larger (smaller) value upward (downward)
by the added in quadratures errors related to the uncertainties in 
$\alpha_s$, $m_b$, $m_c/m_b$, $|V_{tb}V^\star_{ts}/V_{cb}|^2$, and higher 
order electroweak corrections; we do not take into account the variation of 
the scale $\mu_W$. For a given set of the parameters of the MSSM the 
$BR(B\rightarrow X_s\gamma)$ value shown in fig.~\ref{fig:bll5} corresponds to 
the lowest (highest) edge of the resulting band of theoretical predictions, 
if the whole band is above (below) the range allowed by CLEO \cite{CLEO}, and 
to the central point of the overlap of the theoretical and CLEO bands in the 
case such an overlap exists.
 
The results of the scans, shown in fig.~\ref{fig:bll5} demonstrate that the 
the CLEO result for $BR(B\rightarrow X_s\gamma)$ does not eliminate the
points corresponding to the largest values of 
$BR(B^0_{s(d)}\rightarrow\mu^-\mu^+)$ and even does not exhibit any definite 
correlation between the two rates, especially for those points for which
$BR(B^0_s\rightarrow\mu^-\mu^+)$ is very large.
This is mainly due to the fact that 
the (LO) chargino contribution to $BR(B\rightarrow X_s\gamma)$ decreases
with growing sparticle masses whereas its contribution to 
$BR(B^0_{s(d)}\rightarrow\mu^-\mu^+)$ does not. 
\footnote{For the same reason we do not expect strong constraints from the 
$K^0\bar K^0$ or $B^0\bar B^0$ mass differences, which a priori also depend 
on the chargino and stop parameters.}
This allows to hope that
even the full NLO computation of $BR(B\rightarrow X_s\gamma)$ will not
change this picture qualitatively. Moreover, fig.~\ref{fig:bll5}
shows that the present CLEO bound 
$BR(B^0_d\rightarrow\mu^-\mu^+)<6.2\times10^{-7}$ (shown in the upper-right
plot by the vertical solid line)
already puts some weak constraints on the MSSM parameter space in the case
of large $\tan\beta\sim m_t/m_b$ and $M_A\simlt300$ GeV.

{\bf 5. Flavour changing induced by sfermion mass matrices}

Up to now we have assumed that the fermion and sfermion mass matrices are
flavour diagonal in the same basis (the so-called 
super-CKM basis). In this section we consider the effects of nondiagonal 
entries in the sfermion mass matrices. It is customary to parametrize
such nondiagonal entries by the so-called dimensionless mass insertions
\cite{GAGAMASI,MIPORO}:
\begin{eqnarray} 
(\delta^K_{XY})^{IJ}\equiv {(\Delta{\cal M}^2_K)^{IJ}_{XY}\over
\sqrt{({\cal M}^2_K)^{II}_{XX}({\cal M}^2_K)^{JJ}_{YY}}}, ~~~~
\label{eqn:ins}
\end{eqnarray}
where $X,Y=L,R$, $K=u,d,l$, $({\cal M}^2_K)^{II}_{XX}$ are the 
diagonal elements of the $XX$ blocks of the full mass squared matrices,
and  $(\Delta{\cal M}^2_K)^{IJ}_{XY}$ are the off diagonal
entries of the $XY$ blocks.
Most of these insertions are bounded by the existing experimental data
(for review, see refs. \cite{GAGAMASI,MIPORO}). In the case of the 
$B^0 \rightarrow\bar l^Bl^A$
decay the relevant insertions are $\left(\delta^l_{XY}\right)^{AB}$ 
and $\left(\delta^d_{XY}\right)^{I3}$, $I=1,2$.

The first interesting point is to check the effects of the slepton mass
insertions which are the only source of the decays
$B^0 \rightarrow\bar ll^\prime$ (through the box diagrams with 
charginos in the loop). Very strong bounds from nonobservation of 
the transition $\mu\rightarrow e\gamma$ exist only on the 
$(\left(\delta^l_{LR}\right)^{12}$ insertions \cite{GAGAMASI} whereas 
in the case of the 
$B^0 \rightarrow\bar ll^\prime$ decay most important are the 
insertions $\delta^l_{LL}$ on which the bounds are weaker
\footnote{Ref. \cite{GAGAMASI} gives $\left(\delta_{LL}^l\right)^{12} < 0.2 
\left({M_{\tilde\nu}\over0.5 {\rm TeV}}\right)^2$, 
$\left(\delta_{LL}^l\right)^{13} < 700 
\left({M_{\tilde\nu}\over0.5 {\rm TeV}}\right)^2$ and  
$\left(\delta_{LL}^l\right)^{23} < 100 
\left({M_{\tilde\nu}\over0.5 {\rm TeV}}\right)^2$, 
which in most cases are superseded by $\delta_{LL}^l \simlt 1$ 
- the limit in which one of the sneutrinos becomes tachionic.}. 
Taking $m_{C_1} = 100$ GeV, light stop $M_{\tilde{t}_2}\approx 100$ GeV
and adjustnig the slepton sector mass parameters so to keep 
$M_{\tilde l}\simgt90$ GeV, $M_{\tilde\nu}\simgt50$ for 
$\left(\delta_{LL}^l\right)^{13(23)}\approx 0.9$ we get
\begin{eqnarray}
BR(B^0_s\rightarrow\bar{l}l^\prime) &\simlt& 1.6\times 10^{-11}\nonumber \\
BR(B^0_d\to\bar{l}l^\prime) &\simlt& 3.8\times 10^{-13}
\end{eqnarray}
where $ll^\prime = e\tau$ or $\mu\tau$. (The largest rates are obtained
for $|M_2/\mu|\simlt1$ and small stop mixing angle $\theta_t$; the 
result scale approximately as $|\delta_{LL}^l|^2$.) For other parameters 
(heavier stops and charginos) branching fractions for these processes
are, of course, smaller. 

We now discuss the effects of the flavour non-diagonal mass insertions in 
the down-type squark mass matrix and return to the $\bar ll$ finel states. 
The approximate formulae accounting for the
effects of the insertions $\left(\delta^d_{XY}\right)^{IJ}$ are easily
derived in the so-called mass insertion method 
\cite{GAGAMASI,MIPORO} in which fllavour off diagonal  elements 
of the sfermion mass squared matrices are treated as additional interactions.
Usually, the linear appromation in $\left(\delta^d_{XY}\right)^{IJ}$
is sufficient to account for the results obtained with the full 
diagramatic calculation. 
In the case of a nonzero $\left(\delta^d_{XY}\right)^{I3}$ insertion
the dominant contribution is expected to come from the diagrams involving 
gluinos, due to their strong coupling, $g_s=\sqrt{4\pi\alpha_s}$, to quarks
and squarks. 
\footnote{For nonzero $\left(\delta^d_{XY}\right)^{I3}$ insertion  also 
neutralinos contribute; moreover a
nonzero $\left(\delta^d_{LL}\right)^{I3}$ insertion induces, via the CKM
matrix (see e.g. \cite{MIPORO,RO}), nonzero 
$\left(\delta^u_{LL}\right)^{I3}$ insertions which affect in principle
the chargino contribution. Both these effects are small but are taken into 
account in our numerical code.}
(This expectation is confirmed by the numerical computation in which
all one-loop contributions are taken into account.) Since at one-loop there 
are no box diagrams with gluinos we are left  only with the $Z^0$ and 
neutral Higgs boson penguin diagrams. As previously, the latter type of 
penguin diagrams is important only for large values of $\tan\beta$.
Another important remark is that because the change of flavour in the 
gluino diagrams does not originate from the CKM mass matrix, the rates 
of the $B^0_d$ decays need not be suppressed compared to
the rates of the $B^0_s$ decays.

For $\tan\beta$ values not too large, only the $Z^0$ penguin contribution
can be important. Direct computation shows however, that in the 
formula~(\ref{eqn:formv}) terms linear in the mass insertions 
$\left(\delta^d_{LL(RR)}\right)^{I3}$ cancel out
completely between the self energy $\Sigma^V_X$ and the proper vertex
correction $F^V_X$ ($X=L,R$). Because of that, the effects of 
the nonzero $\left(\delta^d_{LL(RR)}\right)^{I3}$ mass insertions, 
even taking into account their quadratic and higher contributions
in gluino exchanges as well as neutralino diagrams, are small for 
$\tan\beta$ values for which the neutral Higgs boson penguin graphs 
are negligible. Larger effects could come only from nonzero 
$\left(\delta^d_{LR}\right)^{I3}$ mass insertions which, however, are
strongly constrained \cite{GAGAMASI,MIPORO}: 
$|\left(\delta^d_{LR}\right)^{13}|<0.07(m_{max}/1 {\rm TeV})$,
$|\left(\delta^d_{LR}\right)^{23}|<0.03(m_{max}/1 {\rm TeV})$
(where $m_{max}\equiv(max(M_{sq},m_{\tilde g})$). Respecting these 
constraints, $BR(B^0_s\rightarrow \mu^-\mu^+)$ 
($BR(B^0_d\rightarrow \mu^-\mu^+)$ ) remains of order $4\times10^{-9}$
($10^{-10})$.

In the case of large $\tan\beta$ we have to compute in the linear 
approximation in the mass insertions both the scalar parts of the 
self energies and the 1PI vertex corrections to the couplings 
$\bar d_Jd_IS^0(P^0)$. For the self energies we get
\begin{eqnarray}
\Sigma_L^S=-{1\over16\pi^2}{8\over3}g_s^2m_{\tilde g}
\left\{\left(\Delta{\cal M}_D^2\right)_{LR}^{IJ}
C_0(m^2_{\tilde g}, M_D^2, M_D^2)\phantom{aaaaaa}\right.\nonumber\\
\left.-\left(\Delta{\cal M}_D^2\right)_{LL}^{IJ}m_b
\left(A_b + \mu\tan\beta\right)D_0(m^2_{\tilde g},M_D^2,M_D^2,M_D^2)\right\}
\label{eqn:sigins}
\end{eqnarray}
where $D_0$ is the standard four-point function
\begin{eqnarray}
D_0(a,b,c,d)={1\over a-b}\left[C_0(a,c,d)-C_0(b,c,d)\right], 
\end{eqnarray}
$m_{\tilde g}$ is the gluino
mass and $M_D$ is the average
mass of the two botom squarks. Similar formula is obtained for 
$\Sigma_R^S$ with the replacement $\left(\Delta{\cal M}_D^2\right)_{LL}
\rightarrow-\left(\Delta{\cal M}_D^2\right)_{RR}$. 

In the same approximation, for the vertex correction $\bar d_Jd_IP^0$
we get
\begin{eqnarray}
F^P_L &=&{1\over16\pi^2}{8\over3}g_s^2m_{\tilde g}\left\{
\frac{1}{v_1}Z_H^{1k}\left(\Delta{\cal M}_D^2\right)_{LR}^{IJ}
  C_0(m^2_{\tilde g}, M_D^2,M_D^2) \right. \nonumber\\ 
&\quad+&\frac{e}{2s_W}\frac{\mu}{M_W}
\left(m_{d_I}\left(\Delta{\cal M}_D^2\right)_{RR}^{IJ}  + 
\left(\Delta{\cal M}_D^2\right)_{LL}^{IJ}m_{d_J}\right)
Z_H^{2k}\tan\beta D_0(m^2_{\tilde g},M_D^2,M_D^2,M_D^2)\label{eqn:fpins}\\ 
&-&\left.Z_H^{1k}\tan\beta \frac{e}{2s_WM_W}A_b 
\left(m_{d_I}\left(\Delta{\cal M}_D^2\right)_{RR}^{IJ}
    + m_{d_J}\left(\Delta{\cal M}_D^2\right)_{LL}^{IJ}\right)
D_0(m^2_{\tilde g},M_D^2,M_D^2,M_D^2)\right\}\nonumber
\end{eqnarray}
where for the three-linear soft term we have used $A_D^{II}\equiv Y_d^I A_b$.
$F^P_R$ is similar, with $\left(\Delta{\cal M}_D^2\right)_{LL}^{IJ}
\leftrightarrow -\left(\Delta{\cal M}_D^2\right)_{RR}^{IJ}$. Combining 
(\ref{eqn:fpins}) and (\ref{eqn:sigins}) according to (\ref{eqn:widehatp}),
we see that $\left(\Delta{\cal M}_D^2\right)_{RL}^{IJ}$ cancels out.
Moreover, since the CP-odd
scalar $A^0$ whose coupling to leptons is enhanced, corresponds to $k=1$
and $Z_H^{21}=\cos\beta\approx0$, the second line in (\ref{eqn:fpins})
is suppressed compared to the third one. Therefore, we can write:
\begin{eqnarray}
\widehat{F}^P_L\approx
-{1\over16\pi^2}{8\over3}g_s^2\frac{e}{2s_W}\frac{m_b}{M_W}\tan^2\beta\mu
\left(\Delta{\cal M}_D^2\right)_{LL}^{IJ}  m_{\tilde g}
 D_0(m^2_{\tilde g},M_D^2,M_D^2,M_D^2)
\end{eqnarray}
where we have retained only $\left(\Delta{\cal M}_D^2\right)_{LL}^{IJ}$
which in $\widehat{F}^P_L$ is multiplied by $m_{d_J}=m_b$ and
neglected $\left(\Delta{\cal M}_D^2\right)_{RR}^{IJ}$ which is 
multiplied by $m_{d_I}$ (in $\widehat{F}^P_R$ it is the other way around).
Similar calculation leads to
\begin{eqnarray}
\widehat{F}^S_L\approx
{1\over16\pi^2}{8\over3}g_s^2\frac{e}{2s_W}\frac{m_b}{M_W}\tan^2\beta\mu
\left(\Delta{\cal M}_D^2\right)_{LL}^{IJ}  m_{\tilde g}
D_0(m^2_{\tilde g},M_D^2,M_D^2,M_D^2)
\end{eqnarray}
Computing the relevant Wilson coefficients we finally find for the
coefficients $a$ and $b$:
\begin{eqnarray}
a~&=&{1\over16\pi^2}{f_B\over2}\frac{m_l}{M_W^2}
\left({e\over s_W}\right)^4\lambda_{tI} \nonumber\\
&\times&
\left[ Y(x_t) -\frac{8}{3}g_s^2 
\left(\frac{s_W}{e}\right)^2
\frac{M_B^2}{M_{A^0}^2}{\left(\delta^d_{LL}\right)^{I3}\over\lambda_{tI}}
\tan^3\beta m_{\tilde g}\mu  
M_D^2 D_0(m^2_{\tilde g},M_D^2,M_D^2,M_D^2 )\right]
\end{eqnarray}
\begin{eqnarray}
b~&=&\frac{1}{16\pi^2}\frac{f_B}{2}\frac{m_l}{M_W^2}
\left({e\over s_W}\right)^4\lambda_{tI} \nonumber\\
&\times&
\left[ -{8\over3}g_s^2 
\left(\frac{s_W}{e}\right)^2
\frac{M_B^2}{M_{A^0}^2}{\left(\delta^d_{LL}\right)^{I3}\over\lambda_{tI}}
\tan^3\beta m_{\tilde g}\mu  
M_D^2 D_0(m^2_{\tilde g},M_D^2,M_D^2,M_D^2 )\right]
\end{eqnarray}

\begin{figure}
\psfig{figure=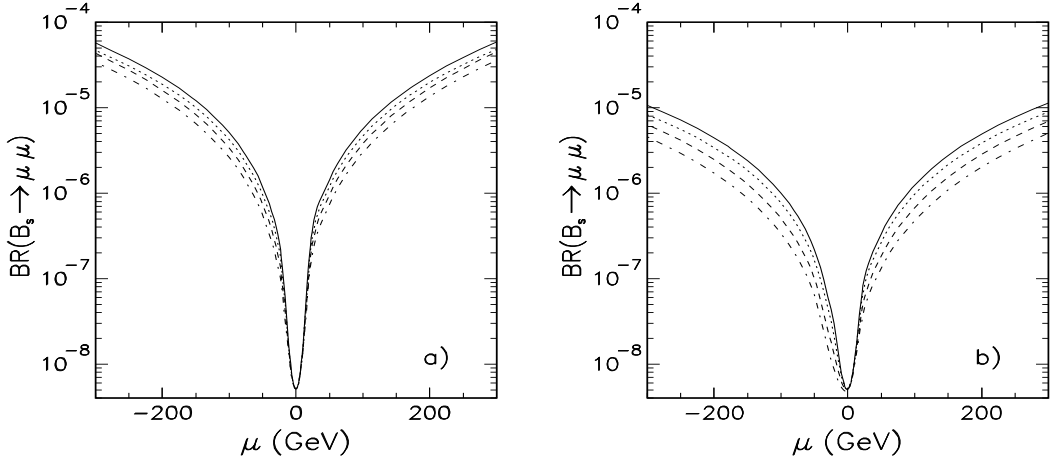,width=15.0cm,height=8.cm} \vspace{1.0truecm}
\caption{$BR(B^0_{s}\rightarrow\mu^-\mu^+)$ for $\tan\beta=50$, $M_A=200$ GeV 
and $(\delta_{LL}^d)^{23}=0.1$
as a function of the $\mu$ parameter. In the left panel
$(m_{\tilde g}, A_t=A_b)$ equals: (300,0) GeV (solid line), (300,250) GeV
(dashed), (800,0) GeV (dotted) and (800,250) GeV (dash-dotted);
$(m^2_Q)_{33}=(500$ GeV$)^2$, $(m^2_U)_{33}=(m^2_D)_{33}=(300$ GeV$)^2$,
$(m^2_X)_{KK}=(600$ GeV$)^2$ for $K\neq3$. In the right panel
$(m_{\tilde g}, A_t=A_b)$ equals: (800,0) GeV (solid line), (800,450) GeV
(dashed), (1500,0) GeV (dotted) and (1500,450) GeV (dash-dotted);
$(m^2_Q)_{33}=(900$ GeV)$^2$, $(m^2_U)_{33}=(m^2_D)_{33}=(700$ GeV)$^2$,
$(m^2_X)_{KK}=1000$ GeV$^2$ for $K\neq3$.}
\label{fig:bll6}
\end{figure}

\noindent in which we have displayed also the SM contribution 
to allow for easy estimate of the magnitude  of the gluino contribution.
It is essential that the dominant effect is due to the $LL$ 
insertion and not the LR one which is much more strongly constrained
\cite{GAGAMASI,MIPORO}. Similarly as in the case of the chargino contribution
through the neutral Higgs boson penguin graphs, also the gluino (and neutralino)
contribution is proportional to $\tan^6\beta$ and does not vanish when 
all SUSY mass parameters are uniformly scaled up (provided the dimensionless
mass insertion is kept fixed).
Figure~\ref{fig:bll6} shows the result of the full diagramatic
computation of the SM and gluino exchange contributions to
$BR(B^0_s\rightarrow\mu^-\mu^+)$ as a function of the $\mu$ parameter
for $(\delta_{LL}^d)^{23}=0.1$ and $\tan\beta=50$, $M_A=200$ GeV. 
The minimum for $\mu=0$ is clearly seen. The gluino contribution 
scales approximately as $|(\delta_{LL}^d)^{23}|^2$.

\begin{figure}
\psfig{figure=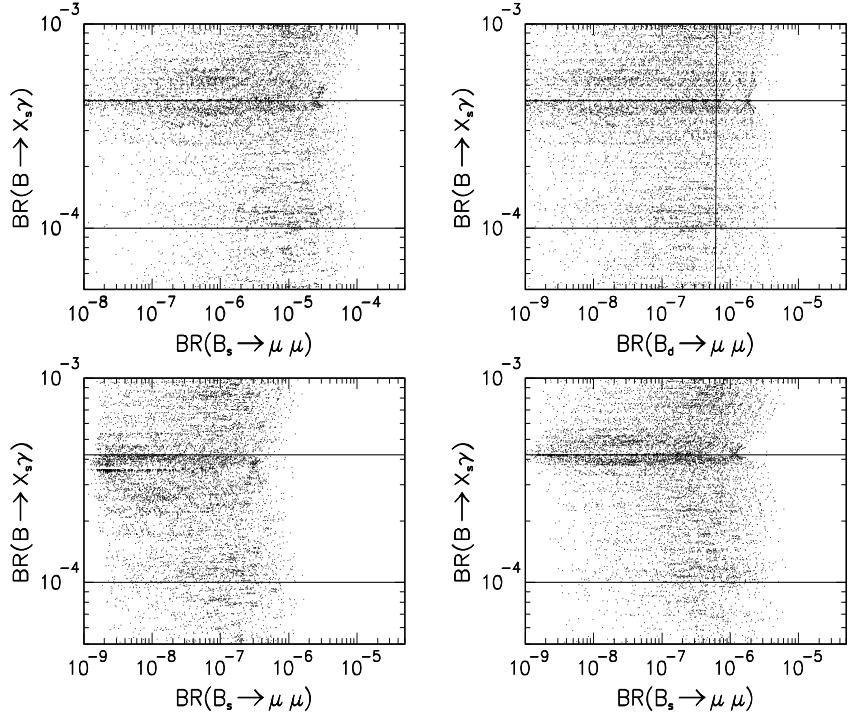,width=12.0cm,height=11.cm} \vspace{1.0truecm}
\caption{$BR(B\rightarrow X_s\gamma)$ versus
$BR(B^0_s\rightarrow\mu^-\mu^+)$ (with $(\delta_{LL}^d)^{23}=0.1$)
and $BR(B^0_d\rightarrow\mu^-\mu^+)$ (with $(\delta_{LL}^d)^{13}=0.05$) for 
$\tan\beta=50$, $M_A=200$ GeV in panels a) and b), $\tan\beta=50$, $M_A=600$ 
GeV in panel c) and $\tan\beta=30$, $M_A=200$ in panel d). Limits from CLEO
on $BR(B\rightarrow X_s\gamma)$ and on 
$BR(B^0_{d}\rightarrow\mu^-\mu^+)$ are also shown by solid lines.}
\label{fig:bll7}
\end{figure}

Figure~\ref{fig:bll7} shows the results of the general scan over the MSSM
parameter space in the form of the scatter plot $BR(B\rightarrow X_s\gamma)$
versus $BR(B^0_s\rightarrow\mu^-\mu^+)$  
for $(\delta_{LL}^d)^{23}=0.1$  and versus
versus $BR(B^0_d\rightarrow\mu^-\mu^+)$  
for $(\delta_{LL}^d)^{13}=0.05$. 
The parameters have been varied in the following ranges: 
$100<m_{C_1}<600$ GeV, $0.1<|M_2/\mu|<5$, $m_{\tilde g}=3M_2$,
$-60^o<\theta_t<60^o$, $A_b = A_t$,
$0.5<M_{\tilde t_2}/m_{C_1}<1.5$,  $1<M_{\tilde t_1}/M_{\tilde t_2}<5$,
$0.25<(m^2_D)_{33}/m^2_{\tilde g}<2.25$. For other entries of the squark mass 
matrices we took $(m^2_X)_{KK}=(m^2_D)_{33}$. All points for which
$M_h<107$ GeV, $\Delta\rho_{sqark}>6\times10^{-4}$ 
(as well as points with too light stops) have been rejected. 
Results for $(\delta_{RR}^d)^{I3}$ are similar. 

In agreement with the bounds given in refs.~\cite{GAGAMASI,MIPORO},
the measured by CLEO \cite{CLEO} $BR(B\rightarrow X_s\gamma)$ does
not constrain the rate of the $BR(B^0_s\rightarrow\mu^-\mu^+)$ decay
(nor does it exhibit any particular correlation with the latter)
and the latter can attain values of the order of $10^{-4}$, respecting
all the relevant phenomenological constraints. As expected,
whenever the gluino contribution is dominant the rates of the 
$B^0_s\rightarrow\mu^-\mu^+$ and $B^0_d\rightarrow\mu^-\mu^+$ decays
are comparable which means that  $BR(B^0_d\rightarrow\mu^-\mu^+)$ can also 
be as large as $10^{-4}$ for $(\delta_{LL}^d)^{13}=0.1$ (in the plot,
we took $(\delta_{LL}^d)^{13}=0.05$ in order to satisfy the 
bound $(\delta_{LL(RR)}^d)^{13}<0.2 (m_{max}/1 {\rm TeV})$ 
\cite{GAGAMASI,MIPORO} for almost all points in the scan; however the biggest 
effects are for $m_{max}$ large for which $(\delta_{LL(RR)}^d)^{13}$ can be 
larger). It follows, that for such values of the MSSM parameters, the current 
CLEO bound, $BR(B^0_d\rightarrow\mu^-\mu^+)<6.2\times10^{-7}$ \cite{WL} puts
constraints on $(\delta_{LL(RR)}^d)^{13}$ which are much stronger than the 
ones given in \cite{GAGAMASI,MIPORO}.

{\bf 6. Conclusions}

We have performed a complete one loop diagramatic calculation of 
the decay rate of the $B^0_{s(d)}$ mesons into charged leptons. 
Both possible sources of the FCNC processe, the CKM mixing matrix 
and the off diagonal entries of the sfermion mass matrices, have 
been considered. For values of $\tan\beta$, for which the neutral 
Higgs boson penguin graphs are negligible the rates of these decays 
in the MSSM remain of the order of the SM prediction. 

Large enhancement of the SM prediction can occur for $\tan\beta\gg1$
provided the additional Higgs bosons predicted by the
MSSM are not too heavy (all the large contributions behave as $1/M^2_A$,
where $M_A$ is the mass of the CP-odd neutral Higgs boson). The 
contribution of the Higgs sector grows like $\tan^4\beta$ and can give 
$BR(B^0_s\rightarrow\mu^-\mu^+)\sim2\times10^{-8}$ for $\tan\beta\sim m_t/m_b$.
Dominant effects of the chargino sector grow as $\tan^6\beta$ and 
depend strongly on the top squark mixing. For 
$\tan\beta\sim m_t/m_b$ and substantial mixing of the top squarks they can 
give $BR(B^0_{s(d)}\rightarrow\mu^-\mu^+)$ up to $5\times10^{-5}$($10^{-6}$)
respecting other phenomenological constraints including the maesurement of 
$BR(B\rightarrow X_s\gamma)$.
Large effects, growing as $\tan^6\beta$ and exhibiting strong
dependence on the $\mu$ parameter, can be also induced
by the off diagonal elements of the down-type squark mass matrix. As we have
shown, $BR(B^0_{s(d)}\rightarrow\mu^-\mu^+)$ is sensitive to the 23 (13)
off-diagonal entries of the LL and RR blocks of these matrices, which are 
not so strongly constrained by $BR(B\rightarrow X_s\gamma)$. For 
$\tan\beta\sim m_t/m_b$ and $M_A\simlt200$ GeV these effects can easily give
$BR(B^0_s\rightarrow\mu^-\mu^+)$ larger than $10^{-4}$. It is also interesting
that even for $BR(B^0_d\rightarrow\mu^-\mu^+)$ these effects can be so large 
that they could exceed the present CLEO limit \cite{WL} which, therefore, 
already now puts constraints on the MSSM parameter space. 

Finally it is important to stress that, both types of effects growing as 
$\tan^6\beta$ do not necessarily decrease as sparticles become heavy.
However, they are sensitive to the mass scale of the extended Higgs sector.
Thus, large deviation from the SM prediction observed in these decays,
apart from being a signal of supersymmetry, would have important implications 
on the Higgs search at the LHC.

\vskip 0.5cm

\noindent {\bf Acknowledgments}
\vskip 0.3cm
\noindent P.H.Ch. would like to thank S. Pokorski for usefull discussion.
His work was partly supported by the Polish State
Committee for Scientific Research grant 2 P03B 052 16 for 1999-2000.

\end{document}